\documentclass[aps,prb,groupedaddress,twocolumn]{revtex4}
\usepackage{graphicx,amsmath,amssymb,bm}
\begin{document}

\title{Tunneling into a Fractional Quantum Hall System and the Infrared Catastrophe}
\author{Kelly R. Patton}
\email[]{kpatton@physnet.uni-hamburg.de}
\affiliation{I. Institut f\"ur Theoretische Physik Universit\"at Hamburg, Hamburg 20355, Germany }
\author{Michael R. Geller}
\email[]{mgeller@uga.edu}
\affiliation{Department of Physics and Astronomy, University of Georgia, Athens, Georiga 30602-2452, USA}
\date{\today}
\begin{abstract}
We calculate the tunneling density of states of a two-dimensional interacting electron gas in a quantizing magnetic field. We show that the observed pseudogap in the density of states can be understood as the result of an infrared catastrophe in a noninteracting electron model. This catastrophe stems from the response of an electronic system to the potential  produced by the abruptly added charge during a tunneling event. Our formalism can be applied at any filling factor without the use of Chern-Simons or composite fermion theory.
\end{abstract}
%\pacs{}
%\keywords{}
\maketitle
\section{Introduction}
Systems where interactions dominate are intrinsically interesting but are also difficult to study theoretically.  A classic example of this is a  quantum Hall system, where the dynamics are mostly, if not entirely, driven by interactions. In such cases, specialized theoretical techniques, such as composite fermion theory, have been developed because standard Fermi liquid theory breaks down in systems with strong correlations. It is these  correlations that leads to the non-trival conductance of the quantum Hall effect.  One might also expect the single-particle spectral properties (i.e.,\ tunneling density of states) of such systems to be influenced by the strong interactions.  In fact, experiments show the low-energy tunneling density of states (DOS) of a quantum Hall system develops a pseudogap. \cite{EisensteinPRL92}  Several theoretical treatments including Chern-Simons theory as well as standard albeit sophisticated diagrammatic approaches have reproduced the pseudogap behavior. \cite{HePRL93, HaussmannPRB96,EricYangPRL93,JohanssonPRB94,KimPRB94}  These specialized approaches are necessary because strong interactions make traditional perturbative methods ineffective. Here we take an alternative approach, focusing on the dynamics of a tunneling event itself, while the interaction between the host particles remain secondary.   

Many systems exhibit a suppression in the DOS near the Fermi energy. In previous work we proposed that the common underlying origin of this suppression is the infrared catastrophe caused by the sudden introduction of a new localized electron into the host system during tunneling. \cite{PattonPRB05,PattonPRB06b,PattonPRB06c} In systems where the accommodation of the new electron is inhibited by dimensionality, an applied perpendicular magnetic field, or disorder  one would expect an infrared catastrophe to occur analogous to that in the x-ray edge problem.\cite{NozieresPR69} In a quantum Hall experiment it is the strong magnetic field that suppresses the recoil of the tunneling electron.   In the limit that the recoil of the added electron is fully suppressed the potential it produces is of the form (assuming the electron is added (tunnels) at $t=0$ and removed at $t_{0}$)
 \begin{equation}
 \label{phixr}
 \phi_{\rm xr}({\bf r},t)=U({\bf r})\Theta(t_{0}-t)\Theta(t),
 \end{equation} 
 where $U({\bf r})$ is the electron-electron interaction.  
  
Contrary to previous approaches,\cite{HePRL93, HaussmannPRB96,EricYangPRL93,JohanssonPRB94,KimPRB94} we show here that the response of a quantum Hall system to potentials of the form Eq.\ (\ref{phixr}) can explain the experimentally observed pseudogap, even while neglecting electron-electron interaction between the host electrons. Previously we applied this idea to the lowest Landau level  (LLL) assuming a delta-function interaction,\cite{PattonPRB05,PattonPRB06b} and we obtained a hard ``gap'' in the DOS.  In this paper we use a more realistic Coulomb form for $U({\bf r})$ and  recover the observed pseudogap.

\section{General Formalism}
\label{general formalism}
This formalism is  generally applicable to a variety of systems and has been discussed in detail in our prior work.\cite{PattonPRB05,PattonPRB06b,PattonPRB06c}  For completeness we restate the general idea here. 
In our previous papers we used the Euclidean time formalism; here, to avoid a difficulty in analytic continuation, we work in real time.  

Starting with a general $D$-dimensional interacting system including a possible external magnetic field the Hamiltonian is taken to be,  
\begin{align}
&H=\sum_{\sigma}\int d^{D}r\ \Psi_{\sigma}^{\dagger}({\bf{r}})\left[\frac{\Pi^{2}}{2m}+v_{0}({\bf{r}})-\mu_{0}\right]\Psi_{\sigma}
({\bf{r}})\nonumber \\  &+\frac{1}{2}\sum_{\sigma \sigma'}\int d^{D}r\ d^{D}r'\ \Psi_{\sigma}^{\dagger}({\bf{r}})\Psi_{\sigma'}^{\dagger}({\bf{r'}})U({\bf{r}}-{\bf{r'}})
\Psi_{\sigma'}({\bf{r'}})\Psi_{\sigma}({\bf{r}}),
\end{align}
where ${\bf{\Pi}}\equiv{\bf{p}}+\frac{e}{c}{\bf{A}}$ and where $v_{0}({\bf{r}})$ is any single-particle potential energy, which may include a periodic lattice potential or disorder or both.  Apart from an additive constant we can write $H$ as $H_{0}+V$, where 

\begin{equation}
\label{hartree ham}
H_{0}=\sum_{\sigma}\int d^{D}r\ \Psi_{\sigma}^{\dagger}({\bf{r}})\left[\frac{\Pi^{2}}{2m}+v({\bf{r}})-\mu\right]\Psi_{\sigma}
({\bf{r}})
\end{equation}
and
\begin{equation}
V=\frac{1}{2}\int d^{D}r\ d^{D}r'\ \delta n({\bf{r}})U({\bf{r}}-{\bf{r'}})\delta n({\bf{r'}}).
\end{equation}
$H_{0}$ is the Hamiltonian in the Hartree approximation.  The single-particle potential $v({\bf {r}})$ includes the Hartree interaction with self-consistent density $n_{0}(\bf{r})$,

\begin{equation}
v({\bf{r}})=v_{0}({\bf{r}})+\int d^{D}r\ U({\bf{r}}-{\bf{r'}})n_{0}({\bf{r'}}),
\end{equation}
where
\begin{equation}
 n_{0}({\bf{r}})=\Big\langle \sum_{\sigma}\Psi_{\sigma}^{\dagger}({\bf{r}})\Psi_{\sigma}({\bf{r}})\Big\rangle_{0}, 
\end{equation}
and the chemical potential in $H_{0}$ has been shifted by $-U(0)/2$. Here $\langle O\rangle_{0}= {\rm Tr}(e^{-\beta H_{0}}O)/{\rm Tr}(e^{-\beta H_{0}})$ denotes an expectation value with respect to the Hartree-level Hamiltonian. In a translationally invariant system the equilibrium density is unaffected by interactions, but in a disordered or inhomogeneous system it will be necessary to distinguish between the approximate Hartree and the exact equilibrium density distributions.  The interaction in (\ref{hartree ham}) is written in terms of the density fluctuation
\begin{equation}
\delta n({\bf r})\equiv\left[ \sum_{\sigma}\Psi_{\sigma}^{\dagger}({\bf{r}})\Psi_{\sigma}({\bf{r}})\right] -n_{0}({\bf r})=n({\bf r})-n_{0}({\bf r}).
\end{equation}

We want to calculate the zero-temperature time-ordered propagator
\begin{equation}
\label{full Green's}
G({\bf r_{\rm f}}\sigma_{\rm f},{\bf r_{\rm i}}\sigma_{\rm i},t_{0})\equiv-i\big\langle T\Psi_{H}({\bf r_{\rm f}}\sigma_{\rm f},t_{0})\Psi_{H}^{\dagger}({\bf r_{\rm i}}\sigma_{\rm i},0)\big\rangle_{H}
\end{equation}
for the interacting system, which can be written (in the interaction representation with respect to $H_{0}$) as
\begin{align}
&G({\bf r_{\rm f}}\sigma_{\rm f},{\bf r_{\rm i}}\sigma_{\rm i},t_{0})=\nonumber \\&-i\frac{\big\langle T\Psi({\bf r_{\rm f}}\sigma_{\rm f},t_{0})\Psi^{\dagger}({\bf r_{\rm i}}\sigma_{\rm i},0)e^{-i\int\limits_{-\infty}^{\infty}dt\ V(t)}\big\rangle_{0}}{\big\langle T e^{-i\int\limits_{-\infty}^{\infty}dt\ V(t)}\big\rangle_{0}}.
\end{align}
Performing a Hubbard-Stratonovich transformation of the form
\begin{equation}
e^{-\frac{i}{2}\int \delta n U \delta n}=\frac{\int D\phi\ e^{\frac{i}{2}\int \phi U^{-1}\phi}e^{-i\int \phi \delta n}}{\int D\phi\ e^{\frac{i}{2}\int \phi U^{-1}\phi}}
\end{equation}
leads to
\begin{equation}
\label{functional int}
G({\bf r_{\rm f}}\sigma_{\rm f},{\bf r_{\rm i}}\sigma_{\rm i},t_{0})={\cal N}\frac{\int D\phi\ e^{\frac{i}{2}\int \phi U^{-1}\phi}g({\bf r_{\rm f}}\sigma_{\rm f},{\bf r_{\rm i}}\sigma_{\rm i},t_{0}|\phi)}{\int D\phi\ e^{\frac{i}{2}\int \phi U^{-1}\phi}},
\end{equation}
where
\begin{align}
\label{little g}
&g({\bf r_{\rm f}}\sigma_{\rm f},{\bf r_{\rm i}}\sigma_{\rm i},t_{0}|\phi)=\nonumber \\ &-i\big\langle T\Psi({\bf r_{\rm f}}\sigma_{\rm f},t_{0})\Psi^{\dagger}({\bf r_{\rm i}}\sigma_{\rm i},0)e^{-i\int dt\int d^{D}r\ \phi({\bf r},t)\delta n({\bf r},t) }\big\rangle_{0}
\end{align}
is a noninteracting correlation function, and ${\cal N}\equiv\big\langle T e^{-i\int\limits_{-\infty}^{\infty}dt\ V(t)}\big\rangle_{0}^{-1}$ is a constant (independent of $t_0$). So far no approximations have been made.  To make any progress one has to determine what the important field configurations in Eq.\ (\ref{functional int}) are and how to integrate them. 

In systems where the recoil of the newly added electron is suppressed by applied fields, disorder, dimensionality or any combination, we propose the important fields are those close to $\phi_{\rm xr}$.  These fields correspond to potentials of  recoilless electrons being added to the system.  
 
If we neglect all fields except $\phi_{\rm xr}$ in Eq.\ (\ref{functional int}),  the so-called  x-ray edge limit. In this limit the fully interacting Green's function Eq.\ (\ref{full Green's}) is given by
\begin{equation}
\label{little g}
G({\bf r_{\rm f}}\sigma_{\rm f},{\bf r_{\rm i}}\sigma_{\rm i},t_{0})\approx{\cal N}g({\bf r_{\rm f}}\sigma_{\rm f},{\bf r_{\rm i}}\sigma_{\rm i},t_{0}|\phi_{\rm xr}).
\end{equation}
Next we define the Green's function
\begin{align}
& G_{\rm xr}({\bf r}\sigma t,{\bf r'}\sigma' t')\equiv\nonumber \\ &-i\frac{\big\langle T\Psi({\bf r}'\sigma ,t)\Psi^{\dagger}({\bf r'}\sigma' ,t')e^{-i\int dt\ d^{D}r\ \phi_{\rm xr}({\bf r},t) n({\bf r},t)}\big\rangle_{0}}{{ Z_{\rm xr}}}
\end{align}
where ${Z_{\rm xr}}(t_{0})\equiv\big\langle Te^{-i\int dtd^{D}r\ \phi_{\rm xr}({\bf r},t)n({\bf r},t)}\big\rangle_{0}$. 
The correlation function in (\ref{little g}) can be written in terms of $ G_{\rm xr}$ and ${ Z_{\rm xr}}$ as 
\begin{align}
\label{g0 rewrite}
&g({\bf r_{\rm f}}\sigma_{\rm f},{\bf r_{\rm i}}\sigma_{\rm i},t_{0}|\phi_{\rm xr})=\nonumber \\ & Z_{\rm xr}(t_{0})\, G_{\rm xr}({\bf r_{\rm f}}\sigma_{\rm f}t_{0},{\bf r}_{\rm i}\sigma_{\rm i}0)\ e^{i\int dt\ d^{D}r\ \phi_{\rm xr}({\bf r},t)n_{0}({\bf r})}.
\end{align}
 Thus the full Green's function in the x-ray edge limit is
 \begin{align}
 \label{x-ray limit}
&G({\bf r_{\rm f}}\sigma_{\rm f},{\bf r_{\rm i}}\sigma_{\rm i},t_{0})=\nonumber \\ & {\cal N}Z_{\rm xr}(t_{0})\, G_{\rm xr}({\bf r_{\rm f}}\sigma_{\rm f}t_{0},{\bf r}_{\rm i}\sigma_{\rm i}0)\ e^{i\int dt\ d^{D}r\ \phi_{\rm xr}({\bf r},t)n_{0}({\bf r})}.
\end{align}
We now turn to finding $G_{\rm xr}$ for  the LLL.

\section{Application to the LLL}
\label{LLL}
Here we apply the formalism of the previous section to the spin-polarized LLL.  A Dyson equation is solved in Section \ref{LLL dyson} for $G_{\rm xr}$.  $Z_{\rm xr}$ is then calculated in Section \ref{eval of Zxr}. These two factors give the DOS in the x-ray edge limit, Eq.\ (\ref{x-ray limit}).  
\subsection{Dyson Equation}
\label{LLL dyson}
 $G_{\rm xr}$ satisfies a Dyson equation given by
\begin{align}
\label{Gxrdyson}
&G_{\rm xr}({\bf r},{\bf r}',t,t')=G_{0}({\bf r},{\bf r}',t,t')\nonumber \\ &+\int dt'' d^{2}r''\, G_{0}({\bf r},{\bf r}'',t-t'')\phi_{\rm xr}({\bf r}'',t'')G_{\rm xr}({\bf r}'',{\bf r}',t'',t')
\end{align}
where $\phi_{\rm xr}({\bf r},t)=U({\bf r})\Theta(t_{0}-t)\Theta(t)$ and $U({\bf r})$ is taken to be the bare Coulomb potential. Choosing ${\bf B}=-B{\bf e}_{z}$ and the symmetric gauge ${\bf A}=(By/2){\bf e}_{x}-(Bx/2){\bf e}_{y}$, the time-ordered zero-temperature noninteracting Green's function projected into the LLL is
\begin{equation}
\label{G0}
G_{0}({\bf r},{\bf r}',t)=i\sum_{m=0}^{\infty}\phi^{}_{m}({\bf r})\phi^{*}_{m}({\bf r}')[\nu-\Theta(t)],
\end{equation}
where
\begin{equation}
\label{LLLeigenfunctions}
\phi_{m}({\bf r})=\frac{r^{m}}{\sqrt{2\pi 2^{m}m!}}e^{-r^{2}/4}e^{im\theta}
\end{equation}
are the noninteracting single-particle eigenfunctions of the LLL with angular momentum quantum number $m$  ranging from zero to infinity and $0<\nu<1$ is the filling factor.
We work in units where  $\hbar=\ell=1$. Here $\ell$ is the magnetic length; $\ell=\sqrt{\hbar c/eB}$.
Note that the Coulomb potential is diagonal in this basis with matrix elements 
\begin{equation}
\label{Coulomb matrix elements}
\int d^{2}r\,\phi^{*}_{m}({\bf r})U({\bf r})\phi^{}_{m'}({\bf r})=e^{2}\frac{\Gamma(m+1/2)}{\sqrt{2}\,m!}\delta_{m,m'}.
\end{equation}

To solve Eq.\ (\ref{Gxrdyson}) we make the ansatz   
\begin{equation}
\label{ansatz}
G_{\rm xr}({\bf r},{\bf r}',t,t')=\sum_{m}a^{}_{m}(t,t')\phi^{}_{m}({\bf r})\phi^{*}_{m}({\bf r}')
\end{equation}
Substituting Eq.\ (\ref{ansatz}) into Eq.\ (\ref{Gxrdyson}) gives
 \begin{widetext} 
\begin{align}
\sum_{m}a^{}_{m}(t,t')\phi^{}_{m}({\bf r})\phi^{*}_{m}({\bf r}')=i\sum_{m}\phi^{}_{m}({\bf r})\phi^{*}_{m}({\bf r}')[\nu-\Theta(t-t')]&+i\sum_{m,l}\phi_{m}({\bf r})\phi^{*}_{l}({\bf r}') \int\limits_{0}^{t_{{0}}}dt''[\nu-\Theta(t-t'')]a_{m}(t'',t')\nonumber \\ &\times\int d^{2}r''\, \phi^{*}_{m}({\bf r}'')U({\bf r}'')\phi^{}_{l}({\bf r}'').
\end{align}
Using Eq.\ (\ref{Coulomb matrix elements}),
\begin{align}
\sum_{m}\phi^{}_{m}({\bf r})\phi^{*}_{m}({\bf r}')\left[a^{}_{m}(t,t')=i[\nu-\Theta(t-t')]+ie^{2}\frac{\Gamma(m+1/2)}{\sqrt{2}\,m!}\int\limits_{0}^{t_{0}}dt''[\nu-\Theta(t-t'')]a_{m}(t'',t')\right].
\end{align}
Thus we need to solve the integral equation
\begin{equation}
\label{Coulomb integral eq.}
a^{}_{m}(t,t')=i[\nu-\Theta(t-t')]+i\lambda_{m}\int\limits_{0}^{t_{0}}dt''[\nu-\Theta(t-t'')]a_{m}(t'',t'),
\end{equation}
where $\lambda_{m}=e^{2}\frac{\Gamma(m+1/2)}{\sqrt{2}\,m!}$.
Eq.\ (\ref{Coulomb integral eq.}) is similar to an integral equation solved in Ref.~\onlinecite{PattonPRB06b}. The solution is 
\begin{equation}
a^{}_{m}(t,t')=i\frac{(\nu-1)\Theta(t-t')+\nu\Theta(t'-t)e^{-i\lambda_{m}t_{0}}}{1-\nu+\nu e^{-i\lambda_{m}t_{0}}}e^{-i\lambda_{m}(t-t')}.
\end{equation}
\end{widetext}
Now using the solution of Eq.\ (\ref{Gxrdyson}) we can evaluate $Z_{\rm xr}$.

\subsection{Evaluation of $Z_{\rm xr}$}
\label{eval of Zxr}

From Appendix \ref{eval of Z} 
\begin{equation}
{Z_{\rm xr}}=e^{M},
\end{equation}
where
\begin{equation}
M=-\int_{0}^{t_{0}}dt\int_0^1d\xi \int d^{D}r\, U({\bf r}) {G}^{\xi}_{\rm xr}({\bf r},{\bf r},t,t^+)
\end{equation}
and
\begin{align}
 {G}_{\rm xr}^{\xi}({\bf r},{\bf r},t,t^+)&=\sum_{m}a^{\xi}_{m}(t,t^{+})|\phi_{m}({\bf r})|^{2}\nonumber \\ &=i\sum_{m}\frac{\nu e^{-i\xi\lambda_{m} t_{0}}}{1-\nu+\nu e^{-i\xi\lambda_{m} t_{0}}}|\phi_{m}({\bf r})|^{2},
\end{align}
where $\xi$ is a rescaling of the interaction strength. 
Performing the time and space integrals gives
\begin{equation}
M=-it_{0}\sum_{m}\lambda_{m}\int \limits_{0}^{1}d \xi\,\frac{\nu e^{-i\xi\lambda_{m} t_{0}}}{1-\nu+\nu e^{-i\xi\lambda_{m} t_{0}}}
\end{equation}
and finally
\begin{equation}
\label{final M}
M=\sum_{m}\ln\left[1-\nu+\nu e^{-i\lambda_{m}t_{0}}\right].
\end{equation}

Because of the long-range action of the Coulomb potential, the infinite sum in Eq.~(\ref{final M}) is divergent. To remove this divergence we use a screened potential with screening length $\alpha$, determined by the experimental setup of Ref.~[\onlinecite{EisensteinPRL92}], where two Hall fluids are separated by a tunneling barrier. Each two-dimensional electron gas (2DEG) screens the other. We choose the distance between the two 2DEGs---typically two to three magnetic lengths---as a value for the screening length. However our final results are not overly sensitive to this value as long as it is in a physically reasonable range. The screening is carried out by introducing a maximum value $M$ of the quantum number $m$ such that
\begin{equation}
\frac{\alpha}{\sqrt{2M}\ell}\ll 1,
\end{equation}
where $\alpha$ is the screening length and $\sqrt{2M}\ell$ is the approximate  radial distance from the origin or tunneling event to the charge density associated with $|\phi_{M}({\bf r})|^{2}$. Thus
\begin{align}
Z_{\rm xr}(t_{0})&={\rm exp}\left\{{\sum_{m}^{M}\ln\left[1-\nu+\nu e^{-i\lambda_{m}t_{0}}\right]}\right\}\nonumber \\ &=\prod_{m=0}^{M}\left[1-\nu+\nu e^{-i\lambda_{m}t_{0}}\right].
\end{align}
\section{ Results}
\begin{figure}
\rotatebox{270}{
\includegraphics[scale=.36]{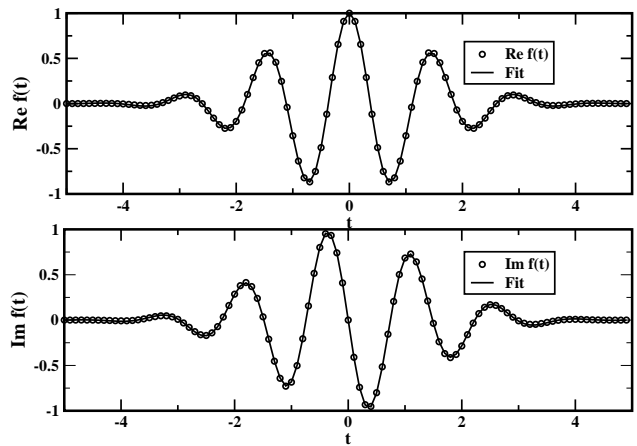}}\vspace{-5mm}
\caption{Example fit of Eq.\ (\ref{fit function}) with $M=50$ and $\nu=.48$ and with $c_{1}\approx 1.0$, $c_{2}\approx -4.2$, and $c_{3}\approx .26$.\label{example fit}}
\end{figure}

The DOS is 
\begin{equation}
N(\omega)=-\frac{1}{\pi}{\rm sgn}(\omega){\rm Im}\,G(\omega)
\end{equation}
where $G(\omega)$ is the Fourier transform of $G(t_{0})$
\begin{equation}
G(\omega)=\int\limits_{-\infty}^{\infty}dt_{0}\, G(t_{0})e^{i\omega t_{0}}.
\end{equation}
In the x-ray edge limit $G(t_{0})$ is given by 
\begin{equation}
G(t_{0})={\cal N}Z_{\rm xr}(t_{0})\, G_{\rm xr}(t_{0})\ e^{i\int dt\ d^{D}r\ \phi_{\rm xr}({\bf r},t)n_{0}({\bf r})}
\end{equation}
or 
\begin{align}
\label{solution}
&G(t_{0})=\nonumber \\ &-{\cal N}\frac{i}{2\pi}(1-\nu)e^{-i\lambda_{0}t_{0}}e^{i\nu e^{2}\alpha t_{0}}\prod_{m=1}^{M}\left[1-\nu+\nu e^{-i\lambda_{m}t_{0}}\right].
\end{align}
The large product of Eq.\ (\ref{solution}) is difficult to handle analytically so we use a numerical fit.  The product can be fit well by a modulated Gaussian with three fit parameters $\{c_{1},c_{2},c_{3}\}$;
\begin{equation}
\label{fit function}
f(t)=\prod_{m=1}^{M}\left[1-\nu+\nu e^{-i\lambda_{m}t}\right]\approx c_{1}e^{ic_{2}t}e^{-c_{3}t^{2}}.
\end{equation}
An example of a fit is shown in Fig.\ \ref{example fit}.

Using the fit function of Eq.\ (\ref{fit function}) the Fourier transform of Eq.\ (\ref{solution}) can be done analytically 
\begin{equation}
G(\omega)=-i{\cal N}\frac{c_{1}}{2\pi}(1-\nu)e^{-\frac{(\omega-\omega_{0})^{2}}{c_{3}}},
\end{equation}
which is a Gaussian centered at $\omega_{0}$ with width $c_{3}^{-1}e^{2}/\ell$.  This leads to a DOS  for $\omega>0$
\begin{equation}
N(\omega)={\cal N}\frac{c_{1}}{2\pi^{2}}(1-\nu)e^{-\frac{(\omega-\omega_{0})^{2}}{c_{3}}}.
\end{equation}
The width of the Gaussian is of the order seen experimentally\cite{EisensteinPRL92} but the energy shift $\omega_{0}$ or more specifically the fit parameter $c_{2}$ can not be accurately determined.  This is because short-time physics enters $Z_{\rm xr}$ through $G_{\rm xr}(t,t^{+})$ and we have solved Eq.\ (\ref{Gxrdyson})  using a long time approximation by including only the LLL in $G_{0}$.  This is a well known problem dating from the original solution of the x-ray edge problem\cite{NozieresPR69} where the threshold energy can not be obtained, nonetheless the exact exponent of the x-ray edge singularity was found. 
\section{Summary}
In this paper we calculated the tunneling DOS in the LLL using the x-ray edge limit approximation and the Coulomb interaction. This limit amounts to the neglection of the recoil of the tunneling electron as well as electron-electron interactions of the host system.  Even in this restrictive limit we are able to  capture the correct physics and obtain the experimentally observed pseudogap.  Although the position of the peak can not be determined from the present calculation the inclusion of higher Landau levels would account for this.

\begin{acknowledgments}
This work was supported by the National Science Foundation under grants DMR-0093217 and CMS-0404031, and by the
German Research Council (DFG) under SFB 668.
\end{acknowledgments}

\appendix
\section{ Calculation  of $Z_{\rm xr}$}
\label{eval of Z}
Here we extend the work of Ref.~[\onlinecite{NozieresPR69}] to include a general $D$-dimensional electron gas with an arbitrary bare interaction, $U({\bf r})$. Using the linked cluster theorem\cite{Mahanbook}
\begin{equation}
Z_{\rm xr}\equiv\big\langle Te^{-i\int dt d^{D}r\ \phi_{\rm xr}({\bf r},t)\hat{n}({\bf r},t)}\big\rangle_{0}=e^{M}
\end{equation}
with
\begin{widetext}
\begin{equation}
M=\sum_{l=1}^{\infty} \frac{(-i)^{l}}{l}\int dt_{1} d^{D}r_{1}\cdots \int dt_{l}d^{D}r_{l}\,\phi_{\rm xr}({\bf r}_{1},t_{1})\cdots\phi_{\rm xr}({\bf r}_{l},t_{l})\left<T\hat{n}({\bf r}_{1},t_{1})\cdots \hat{n}({\bf r}_{l},t_{l})\right>_{{\rm diff., conn.}},
\end{equation}
where only the different and connected terms are retained. Evaluating the exception value
\begin{equation}
M=-\sum_{l=1}^{\infty}\frac{1}{l}\int_{0}^{t_0}dt_{1}\int d^{D}r_{1}\cdots \int_{0}^{t_0}dt_{l}\int d^{D}r_{l}\,U({\bf r}_{1})\cdots U({\bf r}_{l})G_{0}({\bf r}_{1},{\bf r}_{l},t_1,t_l)\cdots G_0({\bf r}_{l},{\bf r}_{1},t_{l},t_{1}^+).
\end{equation}
Changing the summation limit
\begin{equation}
\label{M}
M=-\sum_{l=0}^{\infty}\frac{1}{l+1}\int_{0}^{t_0}dt_{1}\int d^{D}r_{1}\cdots \int_{0}^{t_0}dt_{l+1}\int d^{D}r_{l+1}\,U({\bf r}_{1})\cdots U({\bf r}_{l+1})G_{0}({\bf r}_{1},{\bf r}_{l+1},t_{1},t_{l+1})\cdots G_0({\bf r}_{l+1},{\bf r}_{1},t_{l+1},t_{1}^+).
\end{equation}
Using
\begin{equation}
{G}^{\xi}_{\rm xr}({\bf r},{\bf r}',t,t^+)=G_0({\bf r},{\bf r}',t,t^+)+\xi \int_{0}^{t_0}dt''\int d^{D}r''\, G_0({\bf r},{\bf r}'',t,t'')U({\bf r}'') G^{\xi}_{\rm xr}({\bf r}'',{\bf r}',t'',t^+)
\end{equation}
\end{widetext}
and 
\begin{equation}
\frac{1}{l+1}=\int_{0}^{1}d\xi (\xi)^l
\end{equation}
Eq.\ (\ref{M}) can be written simply as
\begin{equation}
\label{onespin}
M=-\int_{0}^{t_{0}}dt\int_0^1d\xi \int d^{D}r\, U({\bf r})  G^{\xi}_{\rm xr}({\bf r},{\bf r},t,t^+).
\end{equation}
Because we work in the spin-polarized lowest Landau level, Eq.~(\ref{onespin}) is our final needed result. For general systems a sum over spin completes the derivation,
\begin{equation}
M=-\sum_{\sigma}\int_{0}^{t_{0}}dt\int_0^1d\xi \int d^{D}r\, U({\bf r}) G^{\xi}_{\rm xr}({\bf r},{\bf r},t,t^+,\sigma).
\end{equation}

\bibliography{/Users/kpatton/Bibliographies/Master}

%\begin{thebibliography}{99}
%\bibitem{} 
%\end{thebibliography}

\end{document}